\documentclass[12pt]{article}

\usepackage{aaspp4}

\begin{document}

\title{The Influence of Reprocessing in the  Column on
the Light Curves of  Accretion Powered Neutron Stars}

\author{
 Miljenko \v{C}emelji\'{c} and Tomasz Bulik}
 
 \affil{Nicolaus Copernicus Astronomical Center, Bartycka 18, 00-716 Warszawa,
 Poland}
 
\abstract{ Flow of matter onto strongly magnetized neutron stars 
in X-ray binaries
proceeds through accretion funnels that roughly follow geometry of the magnetic field. X-rays originate near surface of the neutron star, and it may happen that the accretion flow passes through the line of sight as the star
rotates. We consider the effects of such accretion flow eclipses
on the X-ray light curves of accretion powered pulsars, and
present a set of X-ray light-curves measured by BATSE for A0535+262 for which this phenomenon is very likely to take place. }

 \section{Introduction}

%What has been done? what not

X-ray binary systems have been observed and analyzed  since
early 1970. They are powered by accretion from a companion star
onto a compact object: neutron star or black hole. In this
paper we concentrate on a subclass of these sources,
namely high mass X-ray binaries. Here, the spin period of a neutron
star is in the range from about less than one second to a couple
hundred of seconds (\cite{WNP-LPH}). Neutron stars are
believed to have strong magnetic fields, in the range of
$10^{12}$~Gauss for a number of reasons. First, lines  have been
discovered in the spectra of some objects at the energies of $5$
to $60$~keV (one object, A0535+262, even has a possible line feature at $110$~keV) and interpreted as cyclotron scattering lines.
Second, the existence of a short pulse period indicates
anisotropy on the surface of the star, which is most likely due to
strong magnetic field.  Since spin periods of these
objects are rather long, typically from ten to even hundreds
seconds, the magnetospheric radius must be large and therefore field is quite strong, or else the star would have
been spun up to much shorter pulse period. The distribution of
magnetic field strength measured by cyclotron lines overlaps with that 
of the radio pulsars magnetic fields measured by the spindown (\cite{Mihara}).

In these systems matter is transferred from the companion star
either through the stellar wind, or also possibly in some
systems through the Roche lobe overflow. An accretion disc may be
formed around a neutron star, however inside the magnetospheric
radius, flow of matter is guided by the magnetic field.  As matter hits the surface of the neutron star, the kinetic energy
in the flow is transformed into radiation and X-rays are
produced. Radiation from the neutron star surface have been
modeled by a number of authors using different approaches to
solve the radiative transfer in a strongly magnetized medium.
Nagel and M\'esz\'aros (1985) have used the finite difference
method, and their approach was later improved  by Alexander and
M\'esz\'aros (1991) who included the effects of higher
harmonics. Bulik et al. (1995) used a similar technique and
solved for the hydrostatic equilibrium with the inclusion of 
the effects of radiation pressure, as well as some temperature
corrections in the atmosphere.  Another method of solving the
radiative transfer - the Monte Carlo  simulations - has been
used in by Wang, Wasserman and Lamb (1993), Araya \& Harding
(1996), however these approaches have only been able to treat
very small optical depths for computational reasons. Recently
this difficulty has been overcome by Isenberg, Wang and Lamb
(1997) who calculated a number of cases with 
optical depth $\tau_T$ up to a few. This model atmospheres have
been used as an input in modeling of the light curves of rotating 
accretion powered pulsars.
The free parameters in these models are the geometry of the
rotation, magnetic field, and the position of the observer with
respect to the system (\cite{Bulik1992,Bulik1995}).

%Absorption in the column as a possible influence on the light curves

Most of the models of the light curves of  accretion powered
pulsars have neglected the interaction with the accretion
column. However the typical density in the column is  $\approx
10^{19}$~cm$^{-3}$ at its base and the direction to the observer
may happen to be  close to the direction along the column. Thus,
for some objects interaction with the matter  in the accretion
column  may be important for the formation of the light curves. 
In section 2 we present a model of emission from a neutron star
and calculate the effects of the accretion column. In section 3
we calculate some representative light curves and compare them
observations of light curves of A0535+262 in a giant outburst in
1994. Finally, we summarize and discuss our results in
section~4.

\section{Radiative Transfer in the Column}

\begin{figure}
\plotone{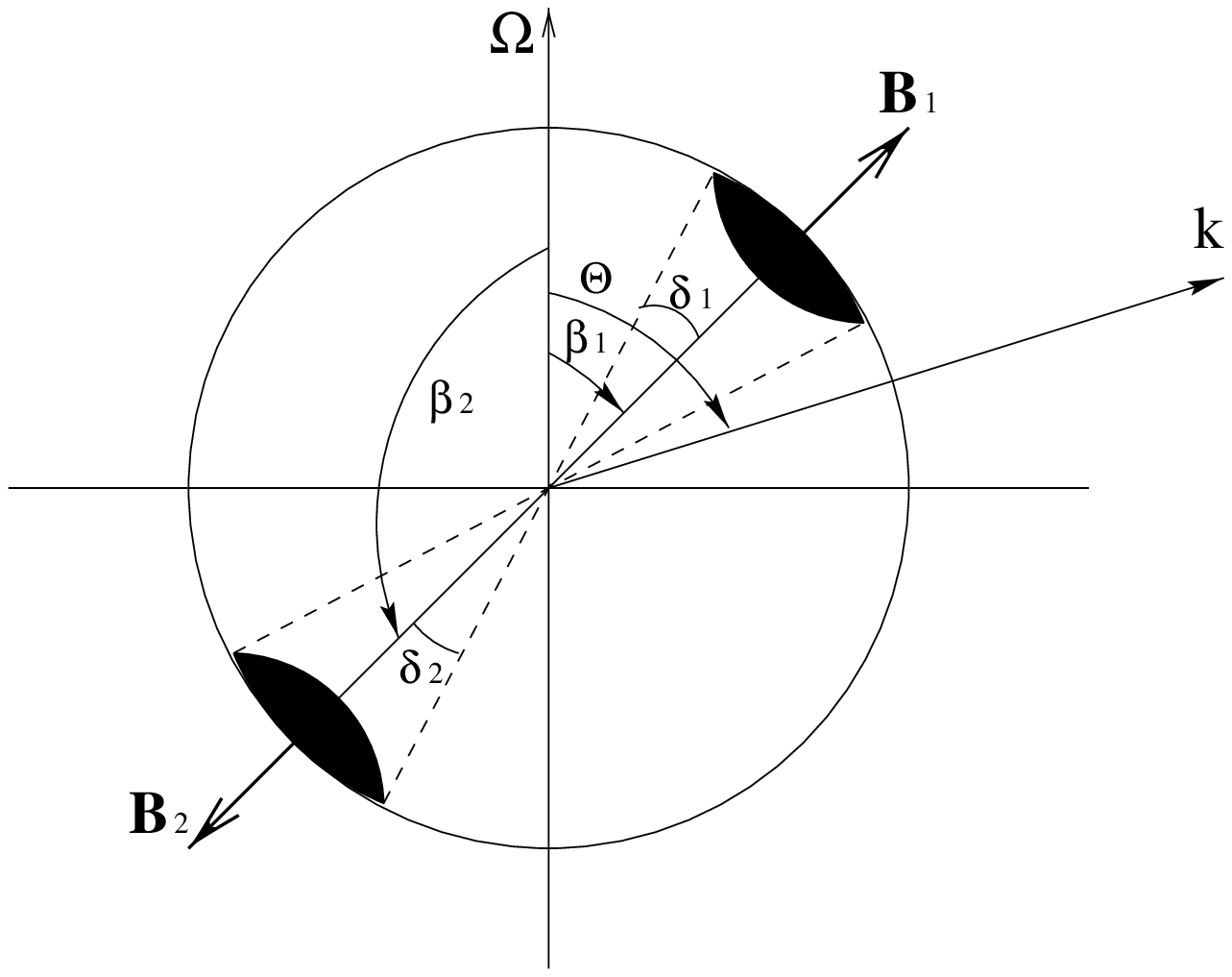}
\caption{Geometry of emission from a rotating  neutron star.}
\label{sketch}
\end{figure}

Our model of emission from a neutron star  follows the models
described elsewhere e.g. Bulik (1995).  The geometry of emission is
shown in Figure~\ref{sketch}. The angle between the direction to the
observer and the neutron star rotation axis is $\Theta$, and the
angle between the magnetic axis and the rotation axis is 
denoted by $\beta$. It should be noted that in general  the two
polar caps do not have to lie opposite one another, as has been
already shown  (\cite{Bulik1992,Bulik1995}).  However, in this
work we will consider only a simplified model  with a single
value of $\beta$. The angular sizes of the accretion caps are
denoted by $\delta$. In this paper we ignore the general
relativistic effects (gravitational light deflection in strong
gravitational field)  which affect the shape of the light curves.
Observationally little is known about the  shape of the
accretion column. It depends on the exact mechanism of how
matter latches onto the magnetic field lines on the edge of the
magnetosphere and also on the geometry of the magnetic field in
the neighborhood of the neutron star.  The accretion flow may
have a geometry of a hollow column, or a fraction of such
(\cite{BaskoSunyaev}), or a filled column. Regardless of the
details, we can say that there must exist a range of lines of
sight for which the observer will look close to the direction
along the accretion column. If the column is bent then there are
directions in which the column passes through the line of sight
to the observer, for some observing directions. In these
directions a significant reprocessing of the underlying
radiation will be very likely to occur and  dips in the
light curve, similar to  eclipses, will take place. 

We use a detailed model of the X-ray emission from
accretion-powered pulsars
(\cite{Bulik1992,Bulik1995}). The radiative transfer
equation is solved using a first order difference scheme and we 
impose the condition of hydrostatic equilibrium on the model atmosphere. 
Calculation of the radiation pressure includes the effects of
resonant scattering and is coupled to the radiative transfer
equation.  The code can incorporate various energy release
profiles in the atmosphere and the final solution satisfies the
condition of radiative equilibrium. We will  not go into further
details of the code, and we will denote the intensity at the
surface of the radiating cap as $I(\omega,\mu)$, ($\omega$ is
the frequency and $\mu$ is the cosine of the angle of
propagation with respect to the magnetic field).

In order to evaluate the relevant radiative processes for the
radiative transfer in the column we need to determine the
physical conditions that prevail there. The density can be
estimated as the free fall density of matter 
\begin{equation}
n = 1.3\times 10^{19} L_{37} R_6^{1/2} A_{10}^{-1}
\left({M_\odot\over M}\right) f^{-1} \mathrm{cm}^{-3}
\end{equation} 
where $L= 10^{37}L_{37} $erg\,s$^{-1}$ is the
luminosity, $R = 10^6 R_6$~cm is the radius of the neutron star,
$A = 10^{10} A_{10}$~cm$^2$ is the area of the accreting cap,
$M$ is the mass of the neutron star and $f$ is the radiative
efficiency.   The density will changes as the matter moves along
the column. Two effects play a role here: on one hand as the
matter accelerates its density decreases. On the other hand
as  matter approaches a neutron star the field lines are
increasingly squeezed and this also increases the density in the
flow.

The accreting material consists primarily of
hydrogen which will be ionized.  We assume that the magnetic
field is dipolar, which should hold pretty well in the region
above the neutron star up to the edge of the magnetosphere. Thus
the value of the field falls down as $B \approx B_\mathrm{surf}
\times (R_*/R)^3$, where $R_*$ is the stellar radius. Note that near the surface the field geometry may well be modified. Since we consider a region only a fraction of
the star radius above the surface,  the magnetic field in the
column is a small factor smaller than  that at the surface. The
ratio of the  free-free absorption cross section is
(e.g. \cite{BulikMiller}) 
\begin{equation} 
{\sigma_{\mathrm{abs}} \over
\sigma_{\mathrm{scat}}} \approx 5.7\times 10^{-7} n_{19} T_6
\omega_{10}^{-3}   
 \end{equation} 
where $n= 10^{19} n_{19}$~cm$^{-3}$ is the density, 
$T = 10^6 T_6$~K is the temperature, and
$\omega = 10 \omega_{10}$~keV is the photon energy.
Thus in the range of interest the opacity 
will be dominated by the scattering.
 The scattering cross section
depends on the polarization of a photon and
the direction of propagation with respect to the magnetic field.
Since most of the emission from a magnetic cap is
in the extraordinary mode, we will only consider
the E-mode cross section, which can be approximated by
\begin{equation} 
{\sigma_{E}\over \sigma_T} \approx
\left({\omega \over \omega +\omega_c}\right)^2\,,
\end{equation}
where $\omega_c = eB/mc$ is the electron cyclotron frequency. 
The optical depth through the column in the E-mode
is
\begin{equation}
\tau = 1.09 \, L_{37} R_6^{1/2} A_{10}^{-1/2} f^{-1}
\, \left( {\omega\over \omega+\omega_c}\right)^2 .
\end{equation}
Here we do not consider the details
of the cross sections in the neighborhood of the electron cyclotron resonance
as far as the reprocessing is concerned. It must be stressed, however,
that the cyclotron resonance is included in the calculation of the
underlying radiation from the accretion cap.
Thus the column is optically thin for most
of the parameter space, especially for strongly magnetized sources. 

As a neutron star rotates we see the accretion caps at 
an angle $\Theta$ which changes as 
\begin{equation}
\cos\theta (\varphi) = \cos\Theta\cos\beta + \sin\Theta\sin\beta\cos\varphi
\label{mu}
\end{equation}
where $\varphi$ is the phase of the rotation. 
The light curves is thus described by
$I(\omega,\varphi) = I(\omega,\mu_1(\phi)) + I(\omega,\mu_2(\varphi))$, 
where  $\mu(\varphi)$ are the cosines of viewing angles for the two accretion 
caps, and are given by equation~(\ref{mu}).
In order
to calculate the effects of the column eclipses we must
know how the optical depth through the  column depends
on the observed phase, i.e. the function $\tau(\varphi)$.
The observed light curves with the inclusion of the
effects of reprocessing in the  column will be
given by
\begin{equation}
I(\omega,\varphi) = I(\omega,\mu_1(\phi))\times\exp\left[-\tau_1(\varphi)\right] 
                  + I(\omega,\mu_2(\varphi))\exp\left[-\tau_2(\varphi)\right] 
\end{equation}
and we ignore the scattered radiation.
We approximate the unknown functions $\tau(\varphi)$
by Gaussian
\begin{equation}
\tau(\varphi) \propto \exp\left[- \left({\varphi-\varphi_0\over
                                        \sigma}\right)^2\right] .
\end{equation}

%Final model - summary ?

Thus the model for the observed light curves is as follows.
Accreted matter is funneled from the disc onto the surface of
the neutron star by the strong magnetic field. We model the 
emission from the magnetic polar cap using the radiative
transfer code  described elsewhere (\cite{Bulik1995}). The
geometry allows us to see either one or two polar caps. However,
the funnel of the accreted matter crosses our line of sight once
every rotational period, and causes the additional strong dip in
the light curve. The parameters of the model are the angles that
describe the geometry and those describing the physical
conditions of the emitting region: the angle between the
rotation axis and the line of sight,  the angle between  the
rotation axis and the position of the magnetic polar cap, the
phase at which the absorption by the accretion  funnel occurs. 

\section{The Light Curves}

In order to visualize the effects of the reprocessing in the
column we calculate a couple of representative light curves. We
assume that the underlying emission from the accretion cap is
characterized by the temperature $kT= 8.1$~keV, and the 
surface magnetic field with cyclotron energy  $\hbar\omega_{C} =
55$~keV. Energy considered in this model is of the 20-30 keV band. In the simulations we ignore the extent of the accretion
caps and also neglect the general relativistic effects.
Such an approach provides only simplified light curves that should
not be used for detailed fits, however here we concentrate on the
reprocessing of radiation in the column.

\begin{figure}
\plottwo{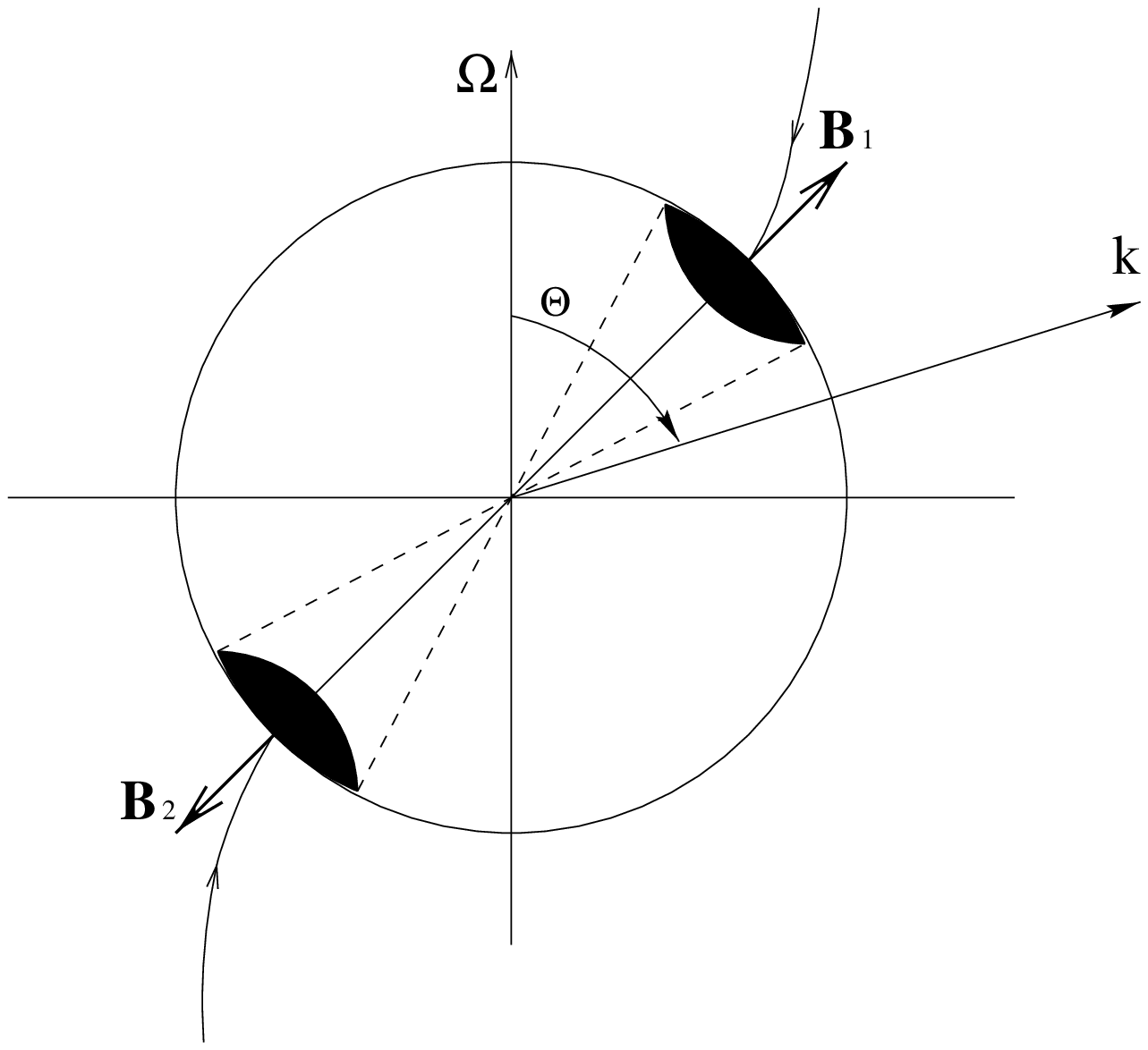}{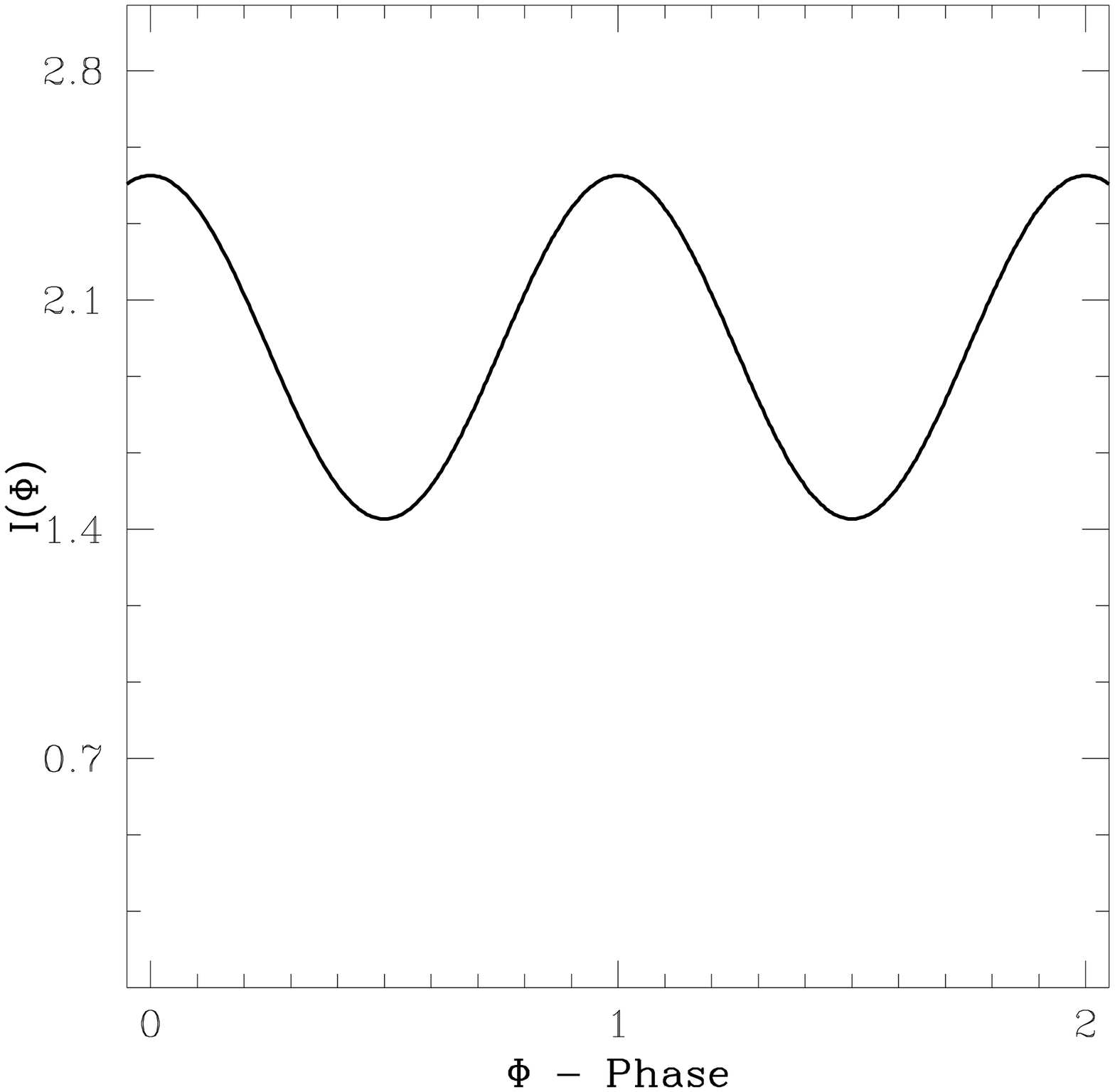}

\plottwo{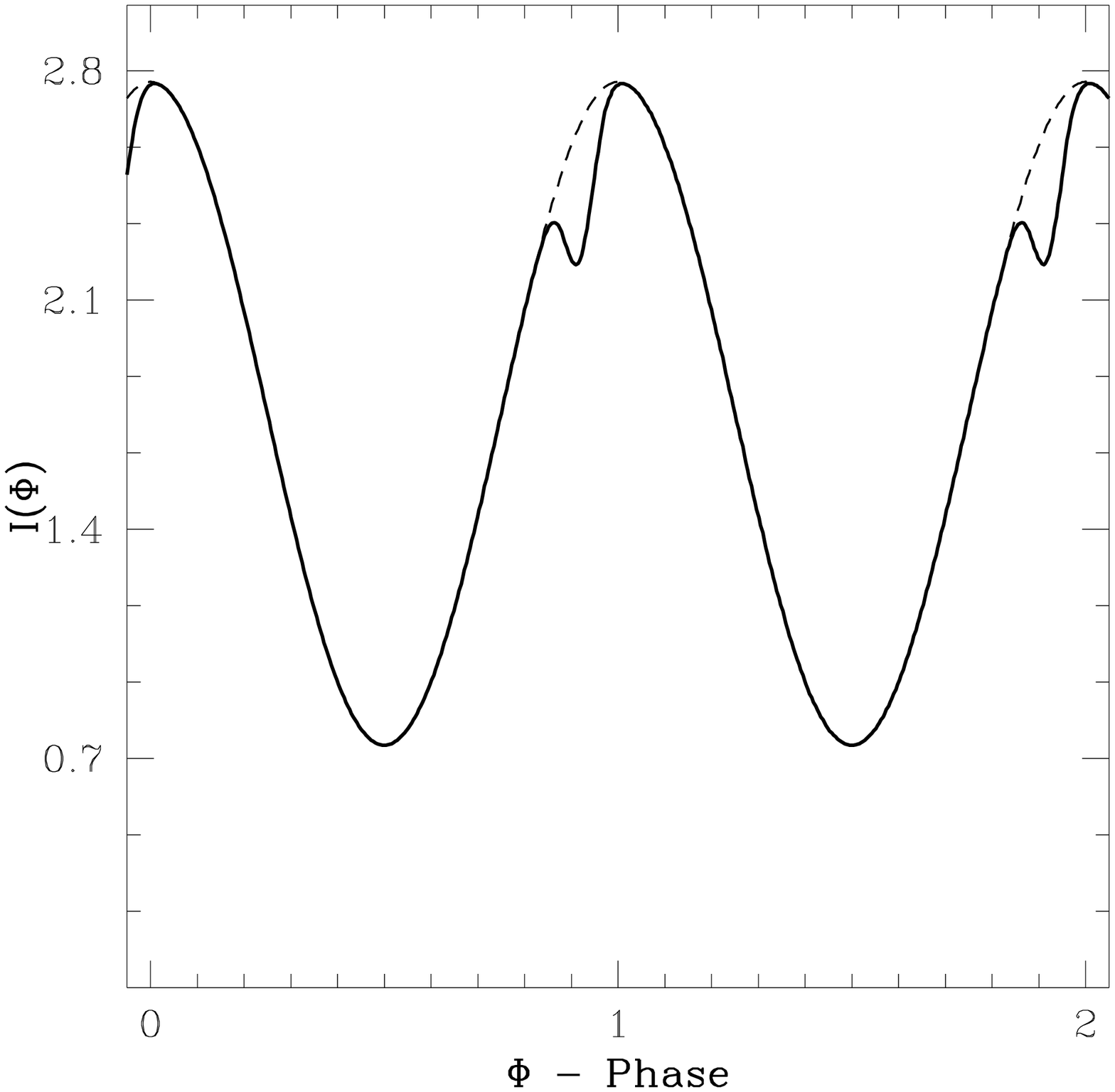}{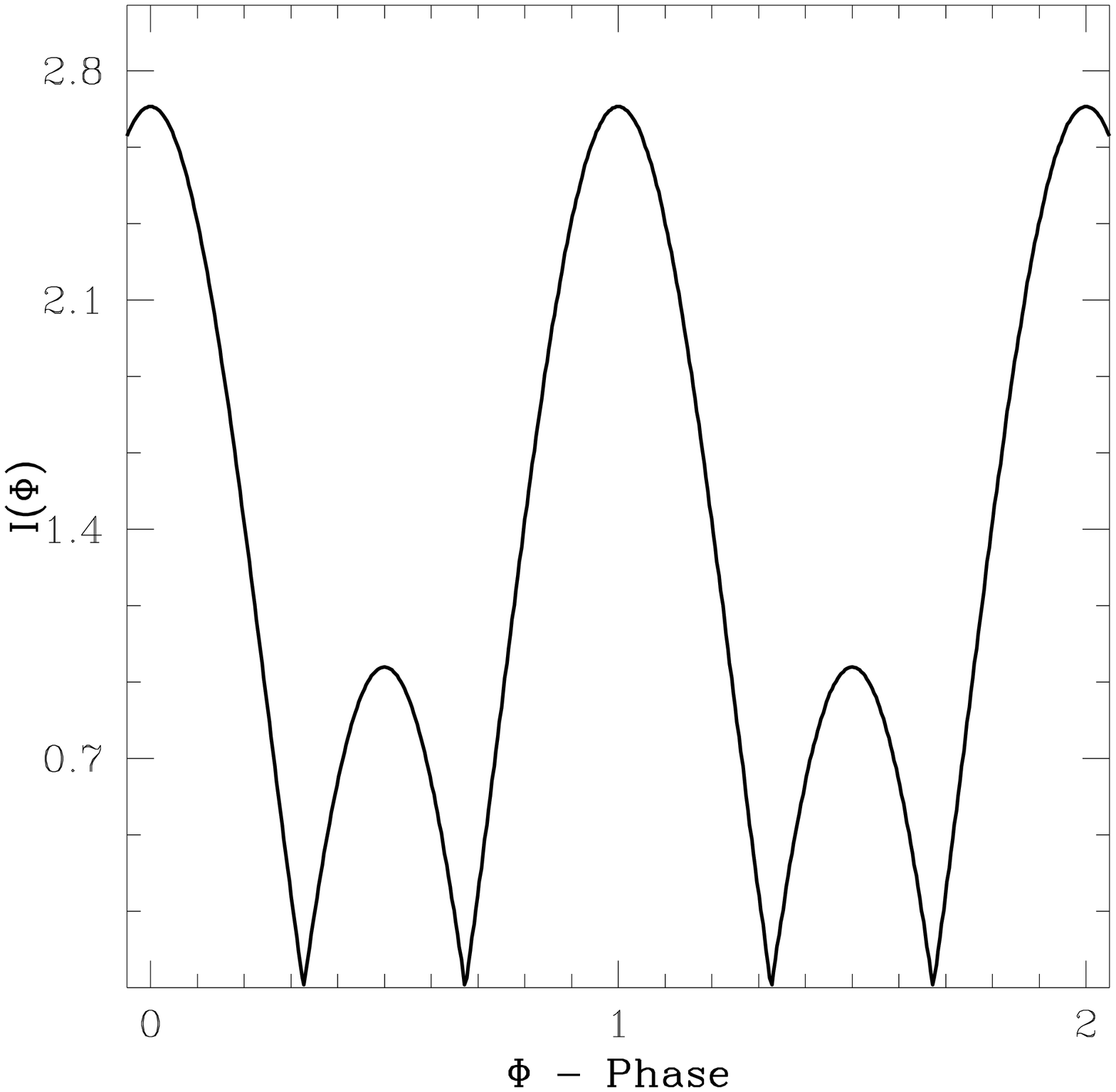}

\caption{The upper left panel shows a sketch of the neutron star
and the accretion flow bending toward the
rotation axis. The panels show light curves for three orientations of the observer in relation to the system,
$\Theta=15^\circ$ (upper right panel), $\Theta=30^\circ$ (lower
left panel), and $\Theta=65^\circ$ (lower right panel).}
\label{bent-up}
\end{figure}

The shape of the light curve depends strongly on the  geometry of
viewing with respect to the system. As a first example we
consider the case when the accretion flow is curved toward the
rotation axis, see Figure~\ref{bent-up}. We assume that the 
angle between both magnetic and rotation axis is $\beta =
45^\circ$, and we present the light-curves for three different 
viewing angles $\Theta=15^\circ, \, 30^\circ$, and $65^\circ$.
In the first case only one accretion cap is seen, and no
reprocessing in the accretion column occurs since the
accretion flow does not cross the line of sight. In the second
case we still see only one accretion cap, however, the line of
sight is crossed by the accretion flow. These results in the
dips in the light curves that produces a double peaked profile.
For clarity we also show with a dashed line a profile that would
be obtained if the accretion flow was ignored. Here  we assumed
that the column density of the accretion flow is
$3\times 10^{24}$cm$^{-2}$, the phase at which the absorption by the accretion funnel occurs $\varphi =330^{\circ}$. It has to be stressed that the depth of
the dips is a very sensitive function of the column depth.
Finally, we show the case when the observing direction is such
that two accretion caps are visible, and the accretion flow does
not cross the line of sight. The light curve has two broad
symmetric  peaks corresponding to seeing two accretion caps. 

\begin{figure}
\plottwo{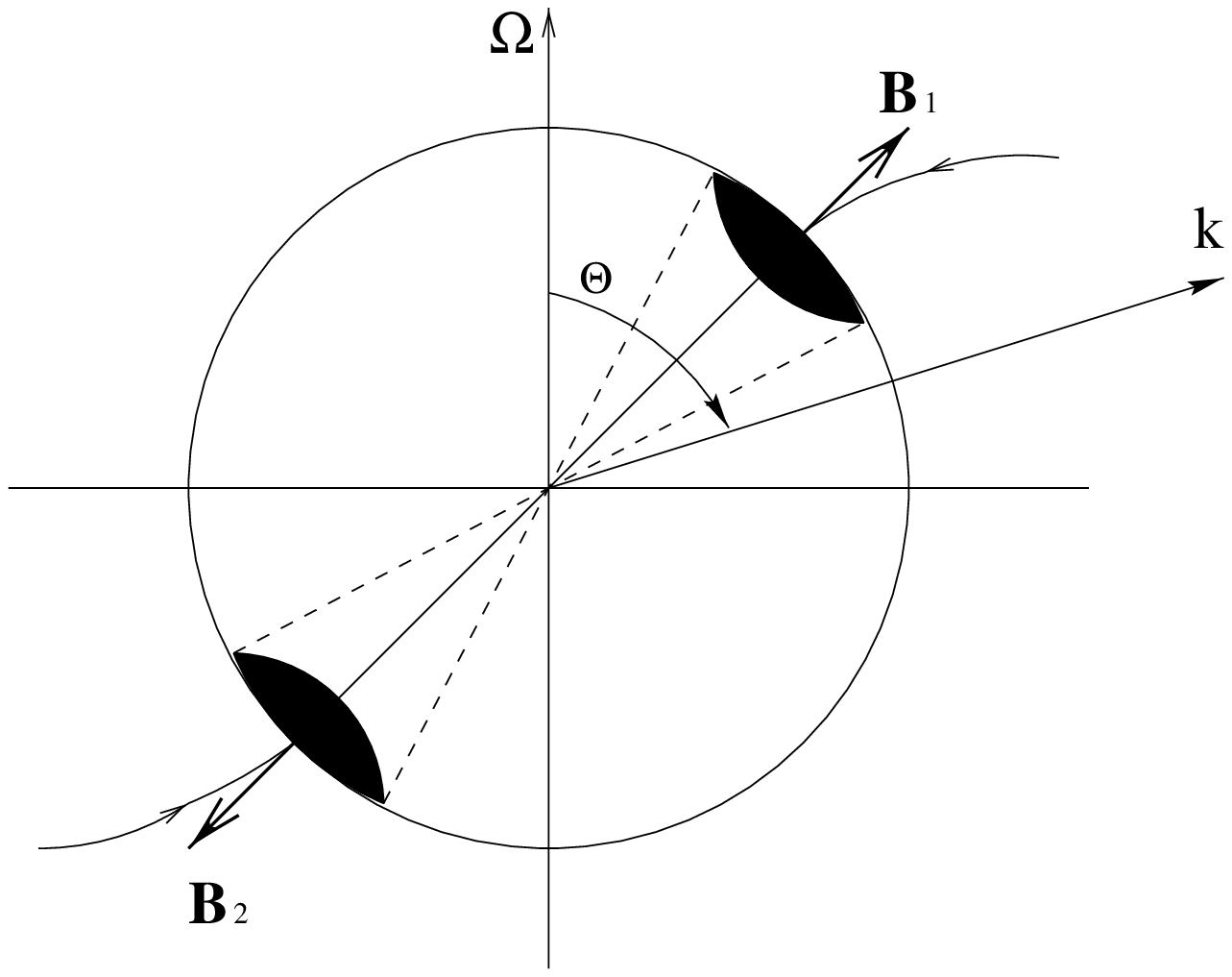}{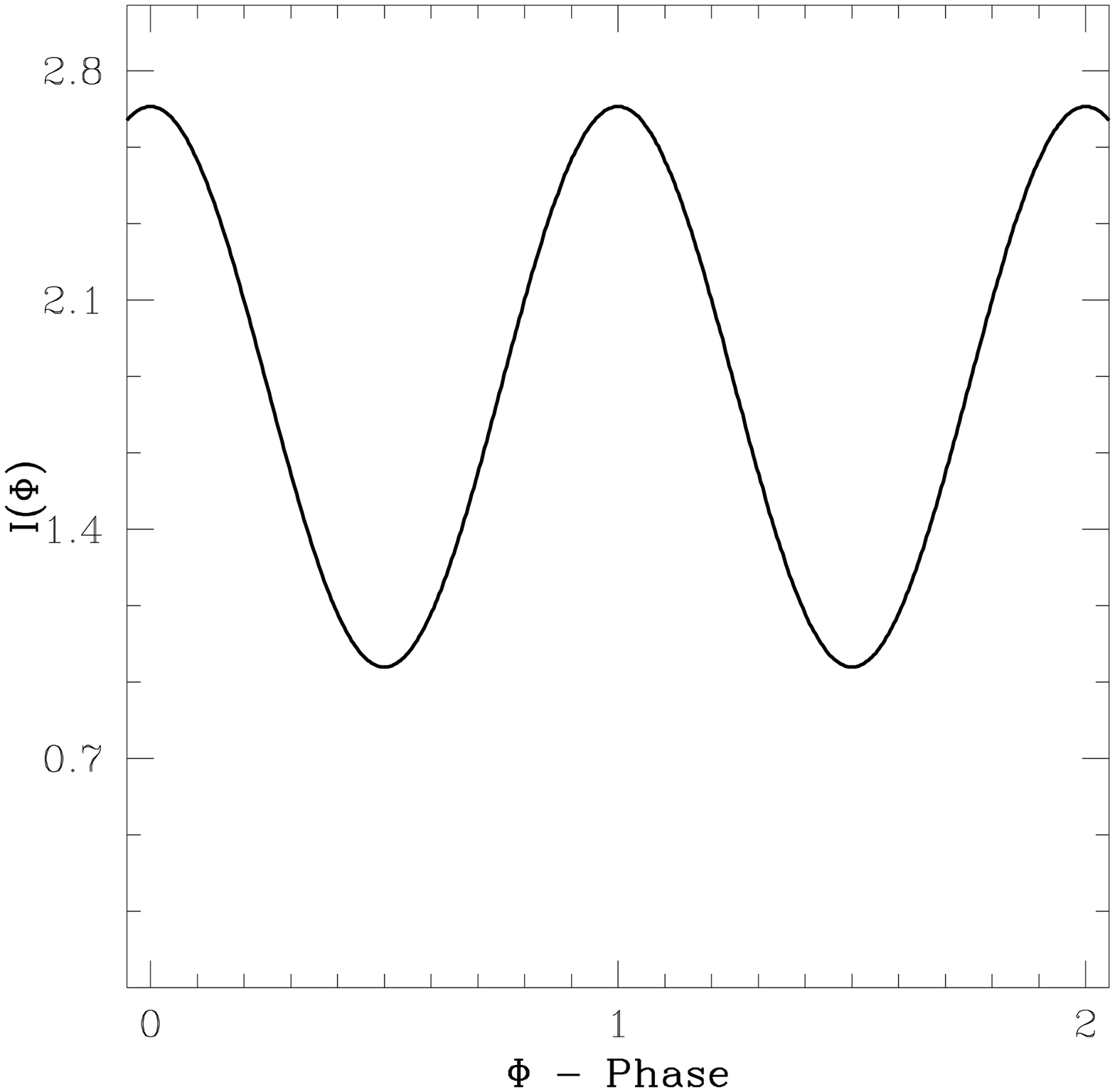}

\plottwo{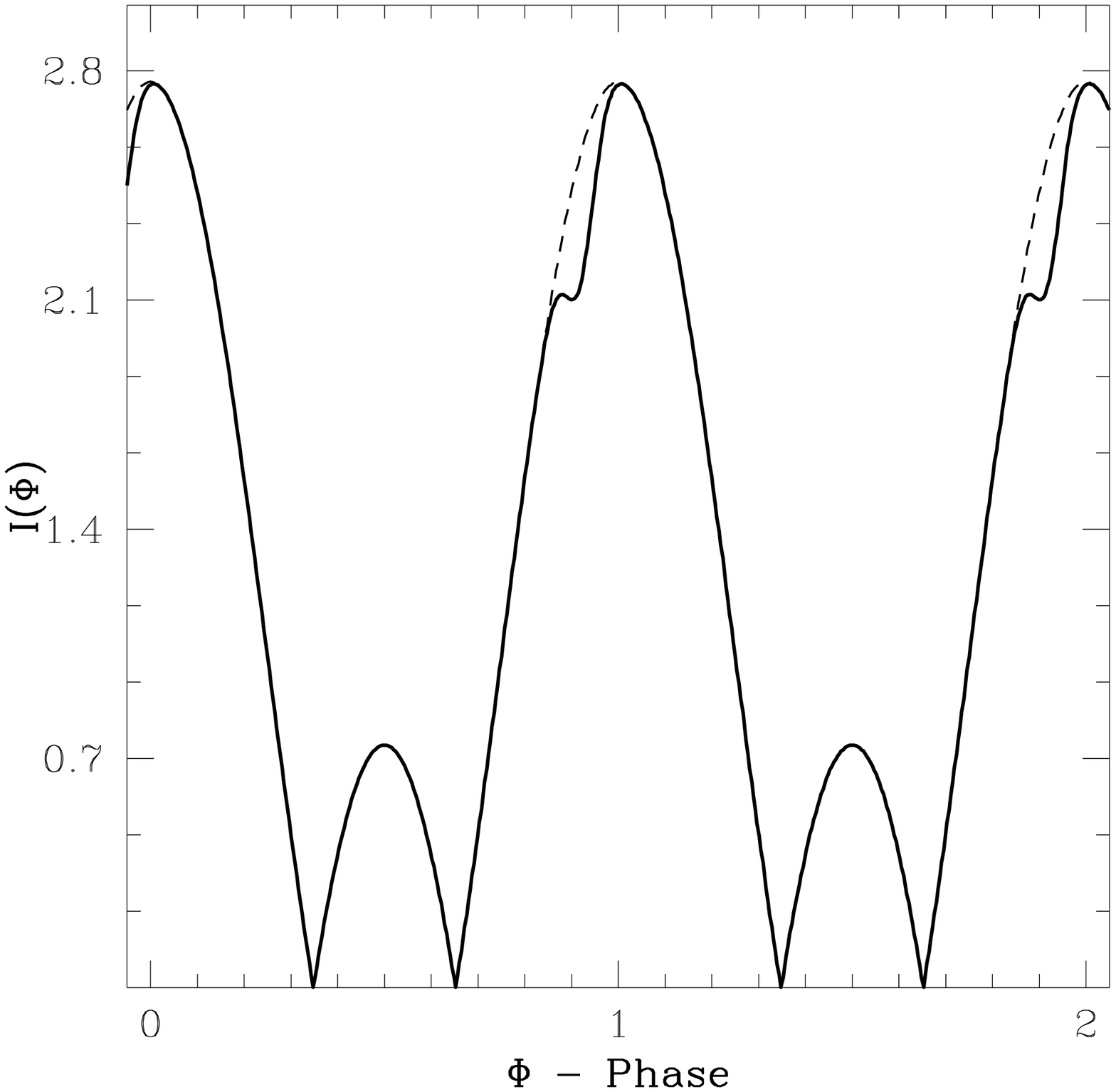}{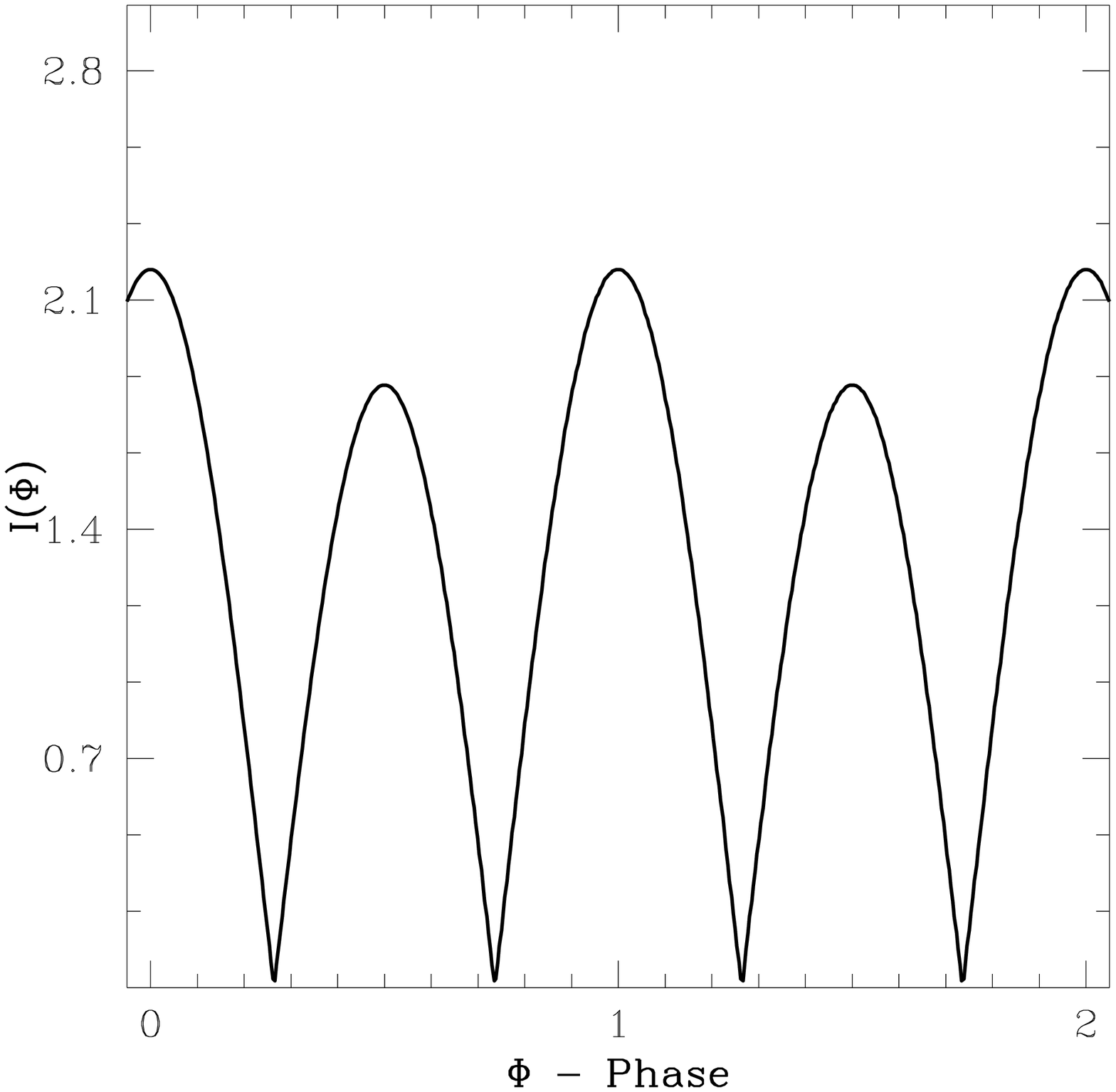}

\caption{The upper left panel shows a sketch of the neutron star
and the accretion flow bending toward the equatorial plane. 
Three panels show light curves 
for three orientations of the observer in relation to the system,
$\Theta=25^\circ$ (upper right panel), $\Theta=60^\circ$ (lower
left panel), and $\Theta=85^\circ$ (lower right panel).}
\label{bent-down}
\end{figure}

The second example set of light curves is shown in
Figure~\ref{bent-down} where we present the case when the
accretion flow is bent towards the equatorial plane. Here  we
also show three light curves for three different viewing angles
$\Theta= 25^\circ,\, 60^\circ$, and $85^\circ$. In the first
case the light curve is similar to the top panel in
Figure~\ref{bent-up}. In the second case, we observe two
accretion caps, however the accretion flow interferes with the
radiation from one of them. This results in a triple peaked
profile, where the larger peak is split because of the scattering
of radiation when the accretion flow passes through the line of
sight. The last case, when the observer is located nearly
perpendicular to the rotation axis, leads to a double peaked
profile, and there are no effects of reprocessing in the
light curve. The position of the  "accretion flow eclipse" dip
with respect to the pulse  center is a measure of bending
of the accretion flow. The larger the difference in phase
between the  pulse center and the dip, the more the flow is bent.

It has to be stressed that the sharp drops in a number of
light curves presented here are not physical. In reality they
will be smoothed because of the gravitational light bending,
as well as the finite extent of the emission regions on the
surface of a neutron star.

\subsection{Light Curves of A0535+262}
\newcommand{\AO}{A0535+262 }

As a very tempting example of a possible application of the
formalism we present the BATSE observations of a binary pulsar
A0535+262. The binary pulsar \AO has been discovered by
{\em Ariel V} 
satellite (\cite{Rosenberg1975,Coe1975}), and the companion star
has been identified as a Be star  (\cite{Liller1975}).
 The binary period has been
estimated  to be $\approx 111$~days from the frequency of X-ray
outbursts (\cite{Nagase1982}). This results was later refined to
$111.38\pm 0.11$~days (\cite{Motch1991}), and the full orbit was
found using the BATSE observations (\cite{Finger1994}). The
rotational period of the neutron star is $103.54$~s
(\cite{Coe1990}).  The intensity of the X-ray outbursts varies 
 and has been classified based on the flux in the 2-10~keV band.
 It is denoted as "giant", when these flux reaches 
above 1~Crab, "normal" when it is  below 
1~Crab,  and "missing" when only the quiescent flux, 5-10~mCrab
is measured.  Only four "giant" outbursts have been observed so
far: in 1975, 1980, 1984, (\cite{Giovanelli1992}), and in 1994.

The X-ray spectrum is thermal and has been modeled as an
exponential with $kT\approx 17$~keV (\cite{Hameury1983}), or a
blackbody or Wien  with $kT\approx 8-9$~keV
(\cite{Frontera1985,DalFiume1988}). Analysis of the HEXE
observation showed a presence of two harmonically spaced lines
(\cite{Kendziorra1992,Kendziorra1994}) at $55$~keV (with low
significance)
and a highly significant line at 
$110$~keV which were interpreted as cyclotron scattering lines.
Recently, OSSE observed \AO (\cite{Grove1995}) and confirmed the
existence of a line at $110$~keV, at the same time
finding no evidence of the line
at  $55$~keV. A further analysis of the data showed that the 
model with $110$~keV line as the fundamental is preferred by the
data (\cite{Araya1996}). Thus, the value of the field   can
either
be  $\approx 5\times 10^{12}$~Gauss or $\approx 10^{13}$~Gauss.

BATSE detected a giant outburst from \AO in February and March
1994 (\cite{Finger1996}),   and triggered a target of opportunity
observation by OSSE (\cite{Grove1995}).  The outbursts was
detectable by BATSE over an interval of 52~days. We should note that BATSE provides a unique opportunity of
almost constant monitoring of  an outburst over a time of next to
two months. Over this time the source changed the flux by a
factor of more than 30. The BATSE experiment can monitor the
intensity of the source using the occultation technique, and also
the light curves folded with the pulse period using the pulsar
mode. A number of such observations have been performed during
the
outburst.

\begin{figure}

\plotone{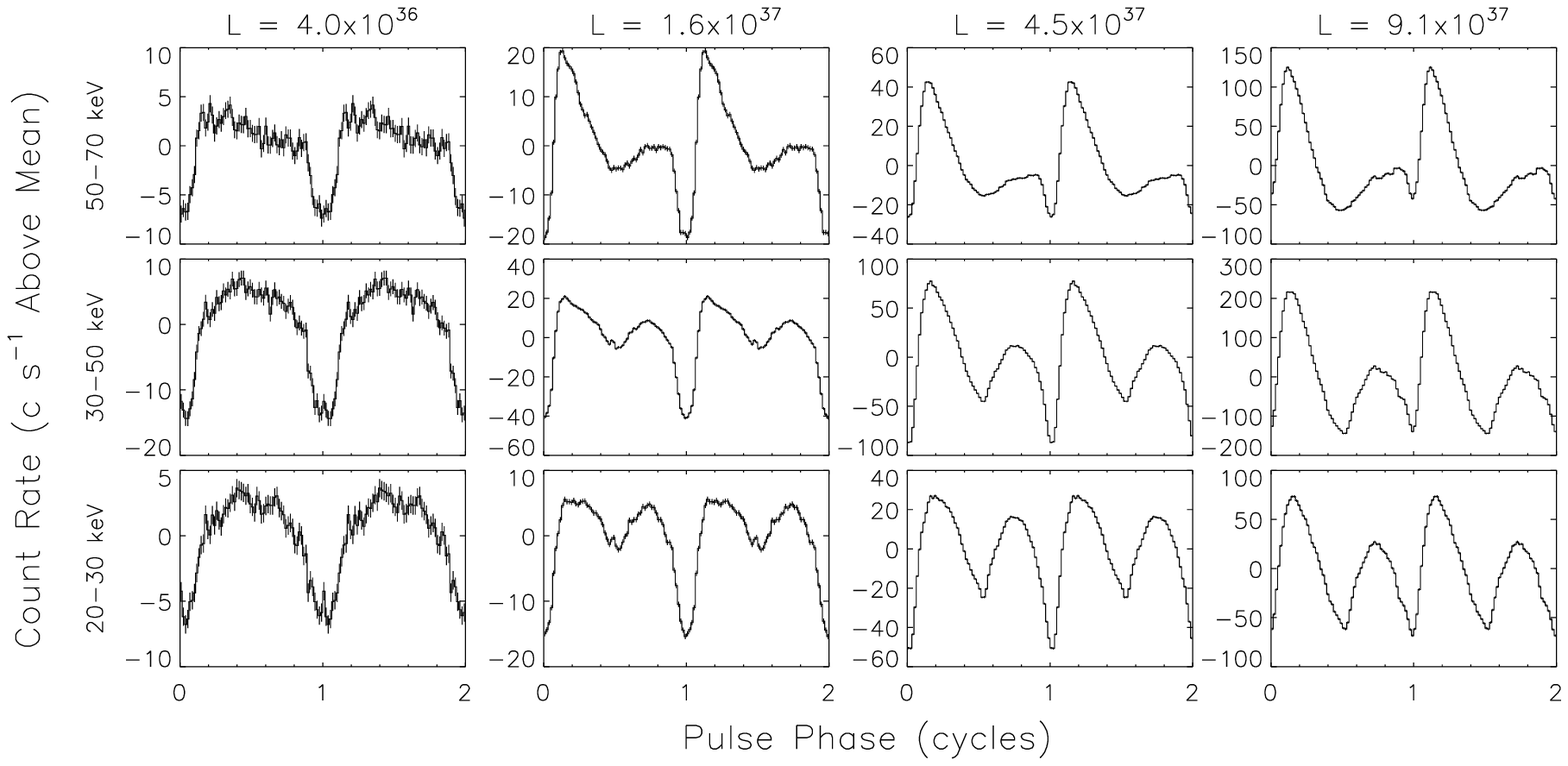}
\vspace{-.5cm}
\plotone{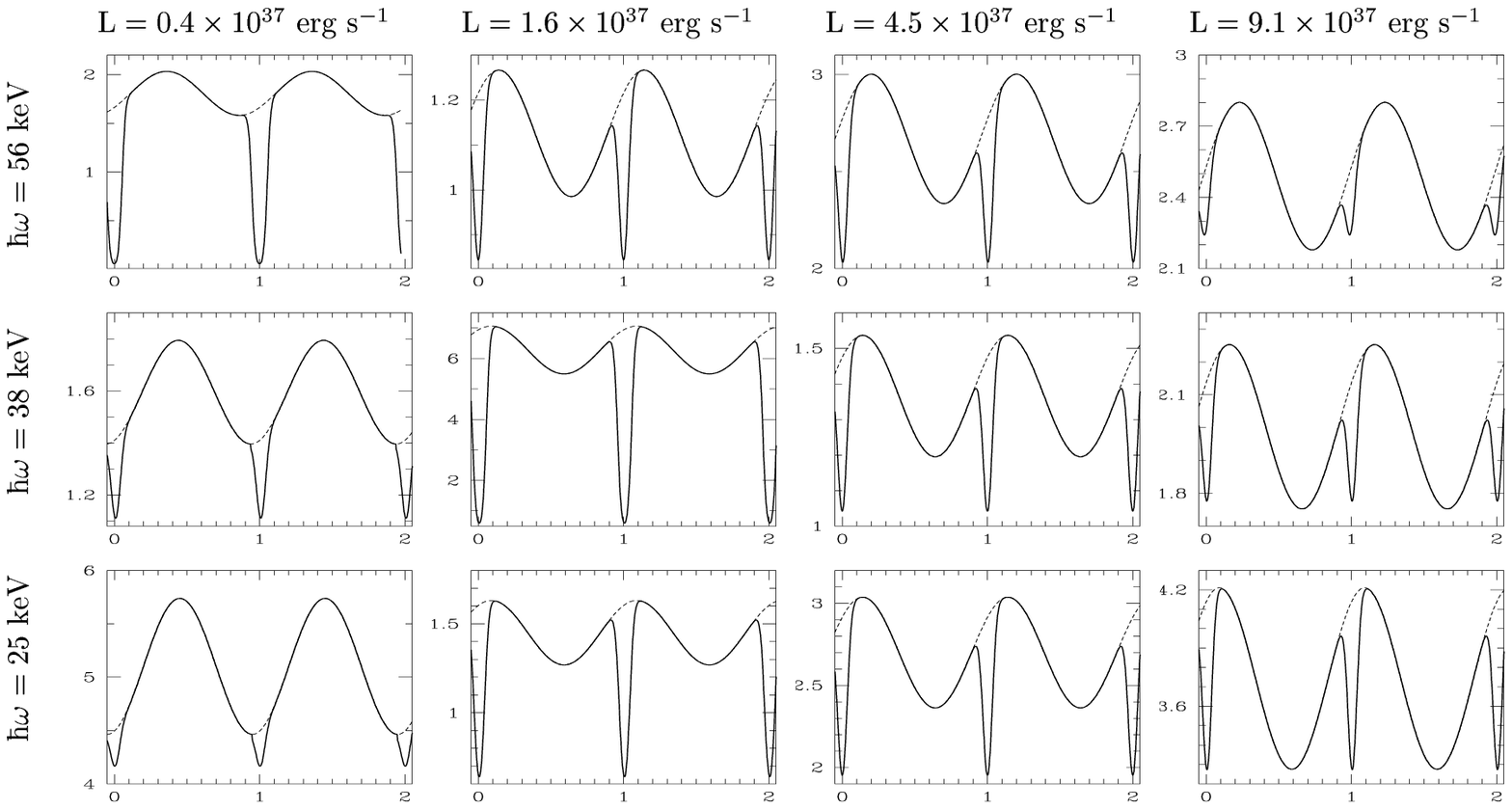}
\vspace{-1cm}
\caption{ The top panel shows light curves of A0535+262 
obtained by BATSE (\protect\cite{Bildsten1997}) folded with the pulse period. 
Columns correspond to four luminosity states when \AO
was observed and are labeled
accordingly. Rows show light curves in three hard X-ray
bands.
The light curves are plotted relative to the mean rate.
The bottom panel shows simulated  light curves of A0535+262. 
Geometry of emission is identical
in the  simulations for all four luminosity states, while 
the accretion rate (or temperature of the emitting region) is varied. The surface
magnetic field
is $\hbar\omega_B = 110$\,keV.
 Dashed lines show the light curves when no absorption by 
 the accretion funnel is taken into account, and solid lines 
 show the light curves with absorption included.}
\label{obssimul}
\end{figure}

The pulse shapes of \AO in different  X-ray bands for four
luminosity states are presented in Figure~\ref{obssimul}. The light
curves  are double peaked except for the lowest luminosity state,
where the light curve is very noisy. The two peaks are divided
by a rather narrow dip with the depth and width varying depending
both on the luminosity and on the spectral band.  The light
curves vary substantially from one spectral band to another as
expected for radiation from a rotating, strongly magnetized
source.

We present a set of simulated light curves for this set of
observations. In this particular calculations we assumed
that  the angle between the rotation axis and the line of sight
is $\theta = 25^\circ$, the angle between the rotation
axis
and  the magnetic axis is $\beta = 15^\circ$. The
temperature of the magnetic cap is estimated assuming a blackbody
radiation
from the area of $10^{10}$\,cm$^2$. We obtain T $=$ 4.5, 6.3, 8.1 and 9.5 keV for the luminosities of L $=$ $0.4\times 10^{37}, 1.6\times 10^{37}, 4.5\times 10^{37}$ and $9.1\times 10^{37}$ erg s$^{-1}$ respectively. We alter the temperature of
the underlying
continuum radiation with the increasing luminosity and we also
change
the density in the column as the luminosity (and consequently
the accretion rate) changes. The results of this qualitative
comparison are shown in Figure~\ref{obssimul}. The detailed fits 
of the model to the data are currently under way, here we want to
stress a qualitative similarity of the simulated and observed
light curves for a wide range of luminosities and spectral ranges.

\section{Summary}

We have analyzed the influence of reprocessing
of radiation when the accretion flow crosses the line of sight in
the accretion powered pulsars.   Some simulated light curves
for different positions of the observer and the source geometry
are presented. We find that in some cases, perhaps
about 10\% of the sources, reprocessing of the radiation in the
accretion column may play a significant role.
There are a few dozen currently known high mass X-ray binaries so
we expect a few sources where such an effect can be seen. 

This opens a very exciting possibility for studying such sources;
the radiation from the accretion cap can serve as a beacon that
shines through the accretion column and allows to probe the
conditions there. Objects that can be observed
for different accretion rates are the best case for such a study. 
As the accretion rate changes the conditions
in the accretion column change and they can be  probed
through analysis of the light curves. Observations of the
underlying continuum, especially if a cyclotron line is present
in their spectrum, provide very strong constraints on the geometry
of the system, e.g. the inclination of the magnetic axis, and
the position of the observer (\cite{Bulik1992,Bulik1995}). 
The spectral analysis of the pulse resolved spectra may yield the
density, temperature, and  possibly
the strength of the magnetic field in the column. One can 
also learn about the geometry of the column by combining the
information of the pulse resolved fits of the  underlying
emission from the cap with the analysis of the accretion flow
dips in the pulses.

A good candidate for such analysis is presented here.
A transient pulsar A0535+262 has been observed by many
satellites. It undergoes outbursts of different strength, and 
 its pulse shape has a dip that is consistent with the
interpretation as  "accretion flow eclipses".

\begin{acknowledgements}
This work has been funded by the KBN grant 2P03D00911.
\end{acknowledgements}

\newpage

\end{document}